\documentclass[preprint,aps,showpacs,amsmath,amssymb,prd,eqsecnum]{revtex4-1}
\usepackage{color}
\usepackage{natbib}
\allowdisplaybreaks

\begin{document}
\bibliographystyle{revtex}

\title{Faddeev-Jackiw quantization and the path integral}

\author{David J. Toms}
\homepage{http://www.staff.ncl.ac.uk/d.j.toms}
\email{david.toms@newcastle.ac.uk}
\affiliation{
School of Mathematics and Statistics,
Newcastle University,
Newcastle upon Tyne, U.K. NE1 7RU}

\date{\today}

\begin{abstract}
The method for quantization of constrained theories that was suggested originally by Faddeev and Jackiw along with later modifications is discussed. The particular emphasis of this paper is to show how it is simple to implement their method within the path integral framework using the natural geometric structure that their method utilizes. The procedure is exemplified with the analysis of two models: a quantum mechanical particle constrained to a surface (of which the hypersphere is a special case), and a quantized Schr\"{o}dinger field interacting with a quantized vector field for both the massive and the massless cases. The results are shown to agree with what is found using the Dirac method for constrained path integrals. We comment on a previous path integral analysis of the Faddeev-Jackiw method. We also discuss why a previous criticism of the Faddeev-Jackiw method is unfounded and why suggested modifications of their method are unnecessary. 
\end{abstract}


\maketitle

\section{Introduction}\label{sec-intro}

The usual method of canonical quantization starts with a Lagrangian $L$ that depends on some generalized coordinates $q^i$ and their time derivatives $\dot{q}^i$. The canonical momentum $p_i$ is identified in the usual way as
\begin{equation}
p_i=\frac{\partial L}{\partial\dot{q}^i},\label{1.1}
\end{equation}
and the Hamiltonian $H$ is defined by a Legendre transformation 
\begin{equation}
H=p_i\dot{q}^i-L,\label{1.2}
\end{equation}
that eliminates the dependence on the generalized velocities $\dot{q}^i$ in favour of the momenta $p_i$. (Summation convention is used here and throughout.) The canonical commutation relations can be written down along with the Heisenberg equations of motion that follow in a familiar way from the classical Poisson bracket relations. The extension to quantum field theory is straightforward. (See \citep{Wentzel} for an early textbook treatment.)

For many theories of interest there are no problems with the implementation of the procedure just described; however, there are important classes of theories where the method fails, for example in electromagnetism. It might not be possible to solve \eqref{1.1} for the velocities $\dot{q}^i$ in terms of $p_i$ and $q^i$. Such theories are said to have a singular Lagrangian and the resulting dynamics is called constrained. A systematic study of constrained dynamics was undertaken by Dirac and his procedure is probably the most widely used. (See \citep{DiracLecturesQM} for Dirac's own review of his method and references to his original work.) Requirements of the method include identifying all of the constraints in the theory, and defining a new bracket, the Dirac bracket, to replace the Poisson bracket. The constraints are classified into two classes: first class, if the matrix formed by the Poisson brackets between constraints is singular, and second class otherwise. The extension of the path integral method to Dirac's theory of constrained systems was presented by Faddeev~\citep{Faddeev1970} and Senjanovic~\citep{Senjanovic}. Various textbook treatments that discuss the Dirac method, as well as other methods for constrained systems, include \citep{Sundermeyer,GitmanTyutin,HenneauxTeitelboim,RotheandRothe}. A clear and succinct account is given in \citep{WeinbergI}.

As an alternative to the Dirac method Faddeev and Jackiw~\citep{FaddeevJackiw} proposed an elegant analysis that can lead to the correct quantum commutation relations without the necessity of the full Dirac machinery. (See also \citep{Jackiwnotears}.) We will give a brief description of their method in the next section. The Faddeev-Jackiw method has received considerable attention since its inception. (A selected set of references includes \citep{barcelos1992symplectic,barcelos1992faddeev,montani1993faddeev,montani1993symplectic,cronstrom1994first,prescod2015extension}.) 
Almost all of the literature, as in the original Faddeev-Jackiw paper~\citep{FaddeevJackiw}, concentrates on the canonical, rather than a path integral, approach. An exception to this is \citep{liao2007path} who discuss how to implement the Faddeev-Jackiw method within the path integral approach. The main purpose of the present paper is to clarify their method slightly by noting that the Jacobian that arises in the path integral measure from the Darboux transformation may be specified in terms of the determinant of the symplectic two-form that arises naturally in the Faddeev-Jackiw method. Furthermore, this identification obviates the need to know what the Darboux transformation actually is as it allows any choice of integration variables to be used. In section~\ref{particle} we will apply this to the case of a particle constrained to an arbitrary surface. This generalizes the hypersphere case considered by \citep{barcelos1992faddeev}. In sections~\ref{Svector} and \ref{Svectormassless} we will consider a field theory example where a Schr\"{o}dinger field is coupled to both a massive and a massless vector field. In both cases we will show how the results agree with the path integrals found from using the more conventional Dirac analysis. We will also comment on some of the literature that is critical of the Faddeev-Jackiw analysis suggesting that it is incomplete and must be modified; our view is that no modifications of the Faddeev-Jackiw analysis are necessary.

\section{Faddeev-Jackiw method}\label{FJmethod}

The starting point of the Faddeev-Jackiw method~\citep{FaddeevJackiw} is the Lagrangian written in first order form. We will also make use of \citep{barcelos1992symplectic,barcelos1992faddeev}. If we call the canonical variables $\xi^\alpha$ then we assume that the Lagrangian takes the general first order form
\begin{equation}
L=A_\alpha(\xi)\dot{\xi}^\alpha+L_v(\xi).\label{2.1}
\end{equation}
The term $L_v(\xi)$ is assumed to contain no time derivatives of the variables $\xi^\alpha$, and it is easy to see that it is the negative of the Hamiltonian. The first term in \eqref{2.1}, that we will refer to as the symplectic part of the Lagrangian, is the main focus of interest. If the original form of the Lagrangian is not first order in time derivatives it is always possible to introduce some auxiliary fields that enable it to be written in the form \eqref{2.1}; usually the normal canonical momenta can be used to do this. The canonical variables $\xi^\alpha$ will then consist of a combination of the original coordinates $q^i$ along with some auxiliary fields and canonical momenta. We will illustrate this in the examples of the next section. The function $A_\alpha(\xi)$ that occurs in \eqref{2.1} is referred to as the canonical one-form.

The equations of motion that follow from \eqref{2.1} (the Euler-Lagrange equations) read
\begin{equation}
F_{\alpha\beta}\,\dot{\xi}^\beta=-\frac{\partial L_v}{\partial\xi^\alpha},\label{2.2}
\end{equation}
where
\begin{equation}
F_{\alpha\beta}=\frac{\partial}{\partial\xi^\alpha}A_\beta - \frac{\partial}{\partial\xi^\beta}A_\alpha.\label{2.3}
\end{equation}
$F_{\alpha\beta}$ defined by \eqref{2.3} is clearly antisymmetric and gives the components of what Faddeev and Jackiw~\citep{FaddeevJackiw} call the symplectic two-form. The Euler-Lagrange equations \eqref{2.2} are invariant under the ``gauge transformation" $\displaystyle{A_\alpha\rightarrow A_\alpha+\frac{\partial}{\partial\xi^\alpha}\Theta}$ for arbitrary $\Theta$. This invariance clearly corresponds to the freedom to add a total time derivative to the Lagrangian which does not affect the equations of motion. 

If $\det F_{\alpha\beta}\ne0$ it follows that we can invert $F_{\alpha\beta}$ in \eqref{2.2} to obtain
\begin{equation}
\dot{\xi}^\alpha=-(F^{-1})^{\alpha\beta}\frac{\partial L_v}{\partial\xi^\beta}.\label{2.4}
\end{equation}
In this case the original Lagrangian is not singular and the usual quantization procedure follows without difficulty. There are no constraints in this case as all of the canonical variables have an evolution equation.

On the other hand if $\det F_{\alpha\beta}=0$ then we cannot invert $F_{\alpha\beta}$ to obtain \eqref{2.4}. In this case some of the canonical variables do not have an evolution equation and the Lagrangian is singular; this means that constraints are present. If $\det F_{\alpha\beta}=0$ then $F_{\alpha\beta}$ necessarily has some zero modes (eigenvectors that correspond to a null eigenvalue). There may be more than one linearly independent zero mode. Call the zero modes $z^\alpha_I$ where $I$ runs over a range that includes all of the linearly independent zero modes that are found for $F_{\alpha\beta}$. By definition,
\begin{equation}
z^\alpha_I\,F_{\alpha\beta}=0.\label{2.5}
\end{equation}
Contraction of both sides of \eqref{2.2} with $z^\alpha_I$ leads to the conditions
\begin{equation}
\Omega_I=z^\alpha_I\,\frac{\partial L_v}{\partial\xi^\alpha}=0.\label{2.6}
\end{equation}
The $\Omega_I$ are the constraints of the theory and they are found by first evaluating the zero modes of the symplectic two-form. 

The next stage in the analysis is to introduce the constraints into the Lagrangian by means of some Lagrange multipliers $\lambda^I$ and to replace \eqref{2.1} with
\begin{equation}
L'=L+\lambda^I\Omega_I.\label{2.7}
\end{equation}
The new term that is added on, $\lambda^I\Omega_I$, is now viewed as part of the symplectic part of the Lagrangian with the canonical variables extended to $(\xi^\alpha,\lambda^I)$. The constraints now give some extra components to the canonical one-form which now has components $(A_\alpha,\Omega_I)$. This is better motivated if instead of adding on $\lambda^I\Omega_I$ to the Lagrangian we instead incorporate the constraints in the form~\citep{barcelos1992faddeev} $\dot{\lambda^I}\Omega_I$. There is no harm in doing this since the Lagrange multipliers are arbitrary and if the constraint holds so must its time derivative. Nevertheless there is no need to do so and we will adopt \eqref{2.7}. From the new canonical one-form we construct the new symplectic two-form $F'_{\alpha\beta}$ as in \eqref{2.3} but now including the new components of the connection one-form as well as the new additions to the canonical variables described above.  We now either find $\det F'_{\alpha\beta}=0$ or else $\det F'_{\alpha\beta}\ne0$. If $\det F'_{\alpha\beta}\ne0$ then the procedure terminates as we can invert $F'_{\alpha\beta}$ as we did in \eqref{2.4}. If $\det F'_{\alpha\beta}=0$ then there must be more zero modes present. We now just repeat the steps \eqref{2.5}--\eqref{2.7} with more Lagrange multipliers added to the set of canonical variables and the canonical one-form extended by taking the extra components to correspond to any new constraints that are found. This procedure is iterated until either an expression is found for the canonical two-form that is not singular(in other words no nontrivial zero modes are present for the canonical two-form), or else the equations \eqref{2.6} hold identically without any new constraints found. In the first case the procedure terminates. In this latter case the symplectic two-form remains singular with the zero modes corresponding to gauge symmetries of the original theory~\cite{montani1993faddeev,montani1993symplectic}. These are dealt with by the usual procedure of adding in gauge conditions which are fixed by additional Lagrange multipliers. Provided that the gauge symmetries are properly dealt with the symplectic two-form is non-singular and the process terminates. The example discussed in Sec.~\ref{Svectormassless} will illustrate this.

Suppose that we have arrived at a non-singular symplectic two-form. We will use $a,b$ as labels here and write the non-singular symplectic two-form as $F_{ab}$ to denote that the canonical variables and the canonical one-form have been extended as described above. The crucial observation made by Faddeev and Jackiw~\citep{FaddeevJackiw} is that Darboux's theorem can be invoked so that the canonical variables $\xi^a$ can be redefined to some new canonical variables $\xi^{\prime a}$ such that the symplectic part of the Lagrangian becomes $\frac{1}{2}\omega_{ab}\xi^{\prime a}\dot{\xi}^{\prime b}$. Here $\omega_{ab}$ is the usual constant symplectic matrix which is necessarily even dimensional and takes the block form
\begin{equation}
\omega_{ab}=\left(
\begin{array}{cc}
0&-I\\
I&0\\
\end{array} \right),\label{2.8}
\end{equation}
with $I$ the identity matrix. Equivalently we can take the symplectic part of the Lagrangian in the standard form $P_\alpha\dot{Q}^\alpha$ where $P_\alpha$ and $Q^\alpha$ are interpreted as canonically conjugate coordinates. (See a nice proof of this in \citep{Jackiwnotears}.) Because $\omega_{ab}$ and $F_{ab}$ just differ by a change of canonical coordinates we have
\begin{equation}
F_{ab}=\frac{\partial\xi^{\prime c}}{\partial\xi^a}\, \frac{\partial\xi^{\prime d}}{\partial\xi^b}\,\omega_{cd}.\label{2.9}
\end{equation}
In practice finding the Darboux transformation might be very difficult, although the constructive proof given by Jackiw~\citep{Jackiwnotears} shows that it exists. A non-trivial example that shows an alternative, but presumably equivalent procedure to obtain the symplectic part of the Lagrangian in the standard canonical form is given by \citep{prescod2015extension}.

However if we adopt the path integral approach it is not necessary to find the Darboux transformation. This is where we differ from \citep{liao2007path}. Because we know that the Darboux transformation reduces the symplectic form to $P_\alpha\dot{Q}^\alpha=\frac{1}{2}\omega_{ab}\xi^{\prime a}\dot{\xi}^{\prime b}$ the formal expression for the path integral measure will be
\begin{equation}
d\mu=\prod_{a}\lbrack d\xi^{\prime a}\rbrack=\left(\prod_{\alpha}\lbrack dP_\alpha\rbrack \right)\left(\prod_{\alpha}\lbrack dQ^\alpha\rbrack \right).\label{2.10}
\end{equation}
If we now perform the transformation from the canonical variables $\xi^{\prime a}$ back to the original set $\xi^a$ the measure will pick up a Jacobian factor:
\begin{equation}
d\mu=\left(\prod_{a}\lbrack d\xi^{a}\rbrack\right)\,J,\label{2.11}
\end{equation}
where
\begin{equation}
J=\det\left(\frac{\partial\xi^{\prime a}}{\partial\xi^b}\right).\label{2.12}
\end{equation}
This is noted by \citep{liao2007path} and the result is left in this form with the implication that to evaluate the measure fully we need to know the explicit Darboux transformation. However this is not the case. The Jacobian is easily evaluated in terms of the symplectic two-form from \eqref{2.9}:
\begin{equation}
J=(\det F_{ab})^{1/2}.\label{2.13}
\end{equation}
The functional measure now becomes simply
\begin{equation}
d\mu=\left(\prod_{a}\lbrack d\xi^{a}\rbrack\right)\,(\det F_{ab})^{1/2}.\label{2.14}
\end{equation}
This means that we can use any set of canonical variables that we like once we have a non-singular two-form. There is no need to know the form of the Darboux transformation explicitly. The only appeal to Darboux's theorem is to justify the form of the path integral measure in \eqref{2.10}. If desired the factor of $(\det F_{ab})^{1/2}$ in the measure can be exponentiated by introducing an integration over real scalar Grassmannian variables.

It is worth noting that Jackiw~\citep{Jackiwnotears} writes down \eqref{2.14} in the case where the original theory is non-singular. The point here is that \eqref{2.14} holds even for a singular theory if the Faddeev-Jackiw procedure in the form advocated by \citep{barcelos1992faddeev} is followed. Because the Faddeev-Jackiw method does not require any real distinction between Dirac's classification into first and second class constraints the measure given in \eqref{2.14} subsumes both; that is, \eqref{2.14} will reproduce both the Faddeev~\citep{Faddeev1970} and Senjanovic~\citep{Senjanovic} form for the path integral. In the next section we will look at three examples and verify that this is the case. We also note that the example contained in \citep{liao2007path} is too simple to show this as it involved no constraints and a unit Jacobian.

\section{Examples}\label{Examples}
\subsection{Particle constrained to a surface}\label{particle}
Consider a particle in $D$-dimensional Euclidean space. Its position can be specified by $D$ Cartesian coordinates $q^i$ with $i=1,\ldots,D$. The usual Lagrangian is
\begin{equation}
L=\frac{1}{2}\,\dot{q}^i\dot{q}_i-V(q),\label{3.1.1}
\end{equation}
where $V(q)$ is some potential. The Kronecker delta can be used to raise and lower indices. This is clearly a non-singular system. Suppose that we now constrain the particle to an arbitrary surface specified by the equation $f(q)=0$. For example, $f(q)=q^iq_i-1$ describes a sphere of dimension $(D-1)$ which is the example studied in \cite{barcelos1992faddeev}. To incorporate the requirement that the particle be constrained to the surface a Lagrange multiplier $\sigma$ is used. We write
\begin{equation}
L=\frac{1}{2}\,\dot{q}^i\dot{q}_i-V(q)+\sigma\,f(q),\label{3.1.2}
\end{equation}
in place of \eqref{3.1.1}.

The first step in the procedure is to write $L$ in the first order form. Define the canonical momentum in the usual way by
\begin{equation}
p_i=\frac{\partial L}{\partial \dot{q}^i}=\dot{q}_i.\label{3.1.3}
\end{equation}
We then have the first order form of the theory given by
\begin{equation}
L=p_i\dot{q}^i +\sigma\,f(q) -\frac{1}{2}p^ip_i-V(q).\label{3.1.4}
\end{equation}
It is convenient to treat $\sigma f(q)$ as part of the symplectic part of the Lagrangian and to take
\begin{equation}
L_v=-\frac{1}{2}p^ip_i-V(q),\label{3.1.5}
\end{equation}
as in \eqref{2.1}. The canonical variables are $\xi^\alpha=(q^i,p_i,\sigma)$. Because there must be an even number of them this means that there must be some constraints present. 

To save introducing cumbersome index sets it is convenient to use the canonical coordinates themselves as component labels. We will define the components of the canonical one-form to be $A_{\xi^\alpha}$ in place of $A_\alpha$, and the symplectic two-form to have components $F_{\xi^\alpha\xi^\beta}$ in place of $F_{\alpha\beta}$. From \eqref{3.1.4} we have
\begin{eqnarray}
A_{q^i}&=&p_i,\label{3.1.6a}\\
A_{p_i}&=&0,\label{3.1.6b}\\
A_{\sigma}&=&f(q).\label{3.1.6c}
\end{eqnarray}
From \eqref{2.3} the components of the symplectic two-form turn out to be, keeping the index order $\xi^\alpha=(q^i,p_i,\sigma)$ and $\xi^\beta=(q^j,p_j,\sigma)$,
\begin{equation}
F_{\xi^\alpha\xi^\beta}=\left(
\begin{array}{ccc}
0&-\delta^{j}_{i}&f_{,i}\\
\delta^{i}_{j}&0&0\\
-f_{,j}&0&0\\
\end{array} \right).\label{3.1.7}
\end{equation} 
Here we have abbreviated $f_{,i}={\partial f(q)}/{\partial q^i}$. It is clear that $\det F_{\xi^\alpha\xi^\beta}=0$ so that the theory is singular with some zero modes. From \eqref{2.5} it can be shown that there is one zero mode given by
\begin{equation}
z^{\xi^\alpha}=(0,f_{,i},1).\label{3.1.8}
\end{equation}
From \eqref{2.6}, noting that ${\partial L}/{\partial\sigma}=0$, we find the constraint
\begin{equation}
\Omega=f_{,i}\,p^i=0.\label{3.1.9}
\end{equation}
This constraint has a simple physical interpretation: the particle momentum is tangent to the surface (its normal derivative vanishes). We could have saved a bit of work by noting that this must hold at the start and enforcing it with another Lagrange multiplier but it is good to show how the procedure works. This constraint also is required to preserve the condition that the original constraint $f(q)=0$ be preserved in time (so that its time derivative vanishes). 

We now modify $L$ by including a Lagrange multiplier $\lambda$ for the constraint $\Omega=0$ in \eqref{3.1.9}:
\begin{equation}
L=p_i\dot{q}^i +\sigma\,f(q)+\lambda\,f_{,i}\,p^i+L_v.\label{3.1.10}
\end{equation}
(We could choose $\dot{\sigma}$ and $\dot{\lambda}$ in place of $\sigma$ and $\lambda$ here if desired, but as we stated above this is not really necessary.) The canonical coordinates may now be extended to include the new Lagrange multiplier: $\xi^\alpha=(q^i,p_i,\sigma,\lambda)$. (We will use the same symbol for $\xi^\alpha$ here as we used before rather than introduce some new name for the variable and some new indexing set; this is an advantage of the notation that we have adopted in which the coordinates themselves are used to label components.) In addition to \eqref{3.1.6a}--\eqref{3.1.6c} we now have an extra component to the canonical one-form:
\begin{equation}
A_\lambda=f_{,i}\,p^i.\label{3.1.10b}
\end{equation}
The components of the canonical two-form now become, with $\xi^\alpha=(q^i,p_i,\sigma,\lambda)$ and $\xi^\beta=(q^j,p_j,\sigma,\lambda)$,
\begin{equation}
F_{\xi^\alpha\xi^\beta}=\left(
\begin{array}{cccc}
0&-\delta^{j}_{i}&f_{,i}&f_{,ik}\,p^k\\
\delta^{i}_{j}&0&0&f^{,i}\\
-f_{,j}&0&0&0\\
-f_{,jk}\,p^k&-f^{,j}&0&0\\
\end{array} \right).\label{3.1.11}
\end{equation} 
This time there are no non-trivial solutions to \eqref{2.5} for the zero modes. (We will verify this by calculating the non-zero determinant of \eqref{3.1.11} below.) We now have a non-singular symplectic two-form and the process terminates. 

In order to compute the determinant of the symplectic two-form in \eqref{3.1.11} we will consider the general block form structure
\begin{equation}
F_{\xi^\alpha\xi^\beta}=\left(
\begin{array}{cc}
A&B\\
C&D\\
\end{array} \right),\label{3.1.12}
\end{equation} 
where $A$ and $D$ are square matrices, but $B$ and $C$ need not be square. The identity (see \cite{ellicott1991geometrical} for example)
\begin{equation}
\left(
\begin{array}{cc}
A&B\\
C&D\\
\end{array} \right)=\left(
\begin{array}{cc}
A&0\\
C&I\\
\end{array} \right)\left(
\begin{array}{ll}
I&A^{-1}B\\
0&D-CA^{-1}B\\
\end{array} \right).\label{3.1.13}
\end{equation}
can be used, assuming that $A^{-1}$ exists, to see that
\begin{equation}
\det\left(
\begin{array}{cc}
A&B\\
C&D\\
\end{array} \right)=(\det\,A)\lbrack\det(D-CA^{-1}B)\rbrack.\label{3.1.14}
\end{equation}
In the case of \eqref{3.1.11} we have
\begin{eqnarray}
A&=&\left(
\begin{array}{cc}
0&-I\\
I&0\\
\end{array} \right),\label{3.1.14a}\\
D&=&\left(
\begin{array}{cc}
0&0\\
0&0\\
\end{array} \right),\label{3.1.14b}\\
B&=&\left(
\begin{array}{cc}
f_{,i}&f_{,ik}\,p^k\\
0&f_{,i}\\
\end{array} \right),\label{3.1.14c}\\
C&=&-B^T.\label{3.1.14d}
\end{eqnarray}
If we now use \eqref{3.1.14} it can be shown that
\begin{equation}
\det\,F_{\xi^\alpha\xi^\beta}=(f_{,i}f^{,i})^2.\label{3.1.15}
\end{equation}

According to \eqref{2.14} the path integral measure is
\begin{equation}
d\mu=\left(\prod_{i}\lbrack dq^i\rbrack\right)\left(\prod_{i}\lbrack dp_i\rbrack\right)\lbrack d\sigma\rbrack\lbrack d\lambda\rbrack\,(f_{,i}f^{,i}). \label{3.1.16}
\end{equation}
The Lagrangian was given in \eqref{3.1.10}. The partition function can be found by performing the integration over the Lagrange multiplier fields $\sigma$ and $\lambda$ to be
\begin{equation}
Z=\int\left(\prod_{i}\lbrack dq^i\rbrack\right)\left(\prod_{i}\lbrack dp_i\rbrack\right)|\nabla f|^2\,\delta(f(q))\,\delta({\mathbf p}\cdot\nabla f) \exp\left\lbrace i\int dt(p_i\dot{q}^i-\frac{1}{2}p_ip^i-V(q) \right\rbrace.\label{3.1.17}
\end{equation}
This formal result agrees precisely with that of Kashiwa~\cite{kashiwa1996general} whose analysis was based on the Dirac formalism. Note that a more precise discretized version of the path integral is given in \cite{kashiwa1996general}.

An alternative form for the partition function $Z$ can be given if we first integrate over $p_i$, then over $\sigma$ and then over $\lambda$ to leave $Z$ in the form of a configuration space path integral:
\begin{equation}
Z=\int\left(\prod_{i}\lbrack dq^i\rbrack\right)|\nabla f|\,\delta(f(q)) \exp\left\lbrace i\int dt(\frac{1}{2}\dot{q}^iG_{ij}\dot{q}^j-V(q) \right\rbrace,\label{3.1.18a}
\end{equation}
where
\begin{equation}
G_{ij}=\delta_{ij}-\frac{f_{,i}\,f_{,j}}{|\nabla f|}.\label{3.1.18b}
\end{equation}
The standard definition for a Gaussian functional integral has been used to obtain this.

\subsection{Schr\"{o}dinger field coupled to a massive vector field}\label{Svector}

We now consider the case of a Schr\"{o}dinger field $\Psi$ coupled to a vector field $B_\mu$ that we will assume is massive. The Lagrangian density is
\begin{eqnarray}
{\mathcal L}&=&\frac{i}{2}\Big(\Psi^\dagger D_0\psi-(D_0\Psi)^\dagger\psi\Big) -\frac{1}{2m}(\mathbf{D}\Psi)^\dagger(\mathbf{D}\Psi)-V\Psi^\dagger\Psi\nonumber\\
&&-\frac{1}{4}W^{\mu\nu}W_{\mu\nu}-\frac{1}{2}m^2\,B^\mu B_\mu.\label{3.2.1}
\end{eqnarray}
Here the gauge covariant derivative is defined by
\begin{equation}
D_\mu=\partial_\mu+ieB_\mu,\label{3.2.2}
\end{equation}
and we use a spacetime metric with signature $(-,+,+,\cdots,+)$. The number of spatial dimensions that we will call $N$ is arbitrary. The Lagrangian is just the integral of \eqref{3.2.1} over the spatial dimensions. The field strength $W_{\mu\nu}$ is the usual one:
\begin{equation}
W_{\mu\nu}=\partial_\mu B_\nu-\partial_\nu B_\mu.\label{3.2.3}
\end{equation}
(The usual notation of $A_\mu$ for the vector field and $F_{\mu\nu}$ for the field strength are eschewed to avoid confusion with the canonical one-form and the symplectic two-form.) The usual local gauge transformation
\begin{eqnarray}
\Psi&\rightarrow& e^{-ie\theta}\Psi,\label{3.2.4a}\\
\Psi^\dagger&\rightarrow& e^{ie\theta}\Psi^\dagger,\label{3.2.4b}\\
B_\mu&\rightarrow&B_\mu+\partial_\mu\theta,\label{3.2.4c}
\end{eqnarray}
is only a symmetry of the theory described by \eqref{3.2.1} if $m^2=0$. We will defer this case until the next section, so that the theory is not locally gauge invariant. 

The Faddeev-Jackiw method requires the Lagrangian to be written in first order form. The Schr\"{o}dinger part that involves $\Psi$ is already first order. The vector field part can be written in first order form by introducing the components of the canonical momenta for the spatial components of the vector field:
\begin{equation}
\pi^i=\dot{B}^i-\partial^i B_0.\label{3.2.5}
\end{equation} 
(Note that with our choice of metric signature $B^i=B_i$ and $B^0=-B_0$.) By separating off the sums over the time and spatial coordinates the first order form of the Lagrangian density reads
\begin{equation}
{\mathcal L}=\pi^i\dot{B}_i+\frac{i}{2}\Psi^\dagger\dot{\Psi}-\frac{i}{2}\dot{\Psi}^\dagger\Psi +{\mathcal L}_v,\label{3.2.6}
\end{equation}
where
\begin{eqnarray}
{\mathcal L}_v&=&-eB_0\Psi^\dagger\Psi-\frac{1}{2m}(\partial_i\Psi^\dagger)\partial^i\Psi-\frac{e^2}{2m}B^iB_i\Psi^\dagger\Psi\nonumber\\
&&+\frac{ie}{2m}B_i\left( \Psi^\dagger\partial^i\Psi- \partial^i\Psi^\dagger\,\Psi\right) - V\Psi^\dagger\Psi\nonumber\\
&&-\frac{1}{2}\pi^i\pi_i-\pi_i\partial^i B_0-\frac{1}{4}W^{ij}W_{ij}+\frac{1}{2}m^2B_0^2 -\frac{1}{2}m^2B^iB_i.\label{3.2.7}
\end{eqnarray}
The canonical variables are $\xi^\alpha=(\Psi,\Psi^\dagger,B_i,\pi^i,B_0)$. There is no problem with having complex coordinates so we do not make the redefinition to real fields as in \citep{liao2007path}. Because there are an odd number of canonical coordinates there must be some constraints. The components of the canonical one-form are
\begin{eqnarray}
A_\Psi&=&\frac{i}{2}\Psi^\dagger,\label{3.2.8a}\\
A_{\Psi^\dagger}&=&-\frac{i}{2}\Psi,\label{3.2.8b}\\
A_{B_i}&=&\pi^i,\label{3.2.8c}\\
A_{\pi^i}&=&0,\label{3.2.8d}\\
A_{B_0}&=&0,\label{3.2.8e}
\end{eqnarray}
Again it is convenient to use the components of the canonical coordinates to label the components of the canonical one-form and the symplectic two-form. The components of the symplectic two-form are defined as in \eqref{2.3} except that for field theory the derivatives must be functional derivatives. Because the formalism uses a constant time hypersurface, only the spatial coordinates will differ for the different components. We have
\begin{equation}
F_{\xi^\alpha\xi^\beta}=\frac{\delta}{\delta\xi^\alpha}\,A_{\xi^\beta} - \frac{\delta}{\delta\xi^\beta}\,A_{\xi^\alpha},\label{3.2.9}
\end{equation} 
analogously to \eqref{2.3}. We choose $\xi^\alpha=(\Psi,\Psi^\dagger,B_i,\pi^i,B_0)$ with all fields evaluated at some spatial coordinate ${\mathbf x}$ and $\xi^\beta=(\Psi^\prime,\Psi^{\prime\dagger},B^{\prime}_{j},\pi^{\prime j},B^{\prime}_{0})$ with all fields evaluated at some spatial coordinate ${\mathbf x}^\prime$. (So for example, $\Psi=\Psi(t,{\mathbf x})$ and $\Psi^\prime=\Psi(t,{\mathbf x}^\prime)$ with similar expressions for the other fields.) It then follows that the components of the symplectic two-form are
\begin{equation}
F_{\xi^\alpha\xi^\beta}=\left(
\begin{array}{ccccc}
0&-i&0&0&0\\
i&0&0&0&0\\
0&0&0&-\delta^{i}_{j}&0\\
0&0&\delta^{j}_{i}&0&0\\
0&0&0&0&0\\
\end{array} \right)\delta({\mathbf x},{\mathbf x}^\prime).\label{3.2.10}
\end{equation}
The symplectic two-form is clearly degenerate with a zero mode $(0,0,0,0,z^{B_0})$ where $z^{B_0}$ is an arbitrary function. This gives rise to a constraint from the consistency condition
\begin{equation}
\frac{\delta}{\delta B_0}\int d^Nx{\mathcal L}_v=0.\label{3.2.11}
\end{equation}
Using \eqref{3.2.7} after an integration by parts of the $\pi^i\partial_i B_0$ term we find the constraint
\begin{equation}
0=-e\Psi^\dagger\Psi+\partial_i\pi^i+m^2\,B_0.\label{3.2.12}
\end{equation}
We now modify the Lagrangian density by adding in the constraint with a Lagrange multiplier $\lambda$, so that
\begin{equation}
{\mathcal L}=\pi^i\dot{B}_i+\frac{i}{2}\Psi^\dagger\dot{\Psi}-\frac{i}{2}\dot{\Psi}^\dagger\Psi +\lambda(\partial_i\pi^i+m^2\,B_0-e\Psi^\dagger\Psi)+{\mathcal L}_v,\label{3.2.13}
\end{equation}
with ${\mathcal L}_v$ unchanged from \eqref{3.2.7}. The new symplectic variables are $\xi^\alpha=(\Psi,\Psi^\dagger,B_i,\pi^i,B_0,\lambda)$ with all fields evaluated at some spatial coordinate ${\mathbf x}$ and $\xi^\beta=(\Psi^\prime,\Psi^{\prime\dagger},B^{\prime}_{j},\pi^{\prime j},B^{\prime}_{0},\lambda^\prime)$ with all fields evaluated at some spatial coordinate ${\mathbf x}^\prime$. There is an additional component to the canonical one-form in addition to those in \eqref{3.2.8a}--\eqref{3.2.8e} which is
\begin{equation}
A_\lambda=\partial_i\pi^i+m^2\,B_0-e\Psi^\dagger\Psi.\label{3.2.14}
\end{equation}
A new row and a new column is added to \eqref{3.2.10} to give
\begin{equation}
F_{\xi^\alpha\xi^\beta}=\left(
\begin{array}{cccccc}
0&-i&0&0&0&-e\Psi^{\prime\dagger}\\
i&0&0&0&0&-e\Psi^\prime\\
0&0&0&-\delta^{i}_{j}&0&0\\
0&0&\delta^{j}_{i}&0&0&\partial^{\prime}_{i}\\
0&0&0&0&0&m^2\\
e\Psi^\dagger&e\Psi&0&-\partial_j&-m^2&0
\end{array} \right)\delta({\mathbf x},{\mathbf x}^\prime).\label{3.2.15}
\end{equation}
There are no zero modes (assuming as we do that $m^2\ne0$.) A careful use of the identity in \eqref{3.1.14} shows that
\begin{equation}
(\det F_{\xi^\alpha\xi^\beta})^{1/2}=\det\lbrack m^2\delta({\mathbf x},{\mathbf x}^\prime)\rbrack.\label{3.2.16}
\end{equation}
This agrees completely with what is found from \citep{Senjanovic} for the massive vector field but where the Dirac procedure for second class constraints is used. 

This substantiates our claim that the functional measure \eqref{2.14} reproduces the result found for a theory that in Dirac language contains second class constraints. (The example of \citep{liao2007path} was too simple to see this as the authors relied on a unit Jacobian.) If the functional integral over $\lambda$ is performed we will end up with a delta function of the constraint given in \eqref{3.2.12}. It is then possible to use this delta function to integrate over the field $B_0$. The fact that $B_0$ is multiplied by $m^2$ in the constraint results in a cancellation of the factor of $m^2$ that arises from the measure factor of \eqref{3.2.16}. The canonical momentum $\pi^i$ can then be integrated out to leave a configuration space path integral that corresponds precisely to what was found by Jackiw~\citep{Jackiwnotears} in a similar example with the Dirac field in place of the Schr\"{o}dinger field and where a canonical analysis was used. Of course Jackiw's~\citep{Jackiwnotears} analysis was much simpler and straightforward than that presented here where he chose to solve the constraint for what we call $B_0$. There is no problem with doing this for this particular example, but there are more complicated situations where the full apparatus described here appears to be necessary~\citep{Tomsinprep}.

\subsection{Schr\"{o}dinger field coupled to a massless vector field}\label{Svectormassless}

We now look at the case where the vector field in \eqref{3.2.1} is massless. From \eqref{3.2.7} the difference now is that the component $B_0$ only enters the Lagrangian linearly. This means that it acts like a simple Lagrange multiplier to enforce the constraint $0=\partial_i\pi^i-e|\Psi|^2$, which is simply the massless limit of \eqref{3.2.12}. Because $B_0$ now occurs linearly we can include it as part of the symplectic part of the Lagrangian density (and we could redefine it to be $\dot{\lambda}$ if we like, but we will not do this). The Lagrangian density in \eqref{3.2.6} can be written as
\begin{equation}
{\mathcal L}=\pi^i\dot{B}_i+\frac{i}{2}\Psi^\dagger\dot{\Psi}-\frac{i}{2}\dot{\Psi}^\dagger\Psi +B_0(\partial_i\pi^i-e\Psi^\dagger\Psi)+{\mathcal L}^{\prime}_v,\label{3.2.17}
\end{equation}
where we define
\begin{equation}
{\mathcal L}^{\prime}_v=\left.{\mathcal L}_v\right|_{B_0=0,m^2=0},\label{3.2.18}
\end{equation}
with ${\mathcal L}_v$ given in \eqref{3.2.7}. The components of the canonical one-form are
\begin{eqnarray}
A_\Psi&=&\frac{i}{2}\Psi^\dagger,\label{3.2.19a}\\
A_{\Psi^\dagger}&=&-\frac{i}{2}\Psi,\label{3.2.19b}\\
A_{B_i}&=&\pi^i,\label{3.2.19c}\\
A_{\pi^i}&=&0,\label{3.2.19d}\\
A_{B_0}&=&\partial_i\pi^i-e\Psi^\dagger\Psi.\label{3.2.19e}
\end{eqnarray}
The terms in \eqref{3.2.19a}--\eqref{3.2.19d} are the same as those in \eqref{3.2.8a}--\eqref{3.2.8d} but now \eqref{3.2.19e} replaces \eqref{3.2.8e}. The net effect of this is the same as if we had dropped $B_0$ altogether from the original formalism, adopted the massless limit of \eqref{3.2.13} with \eqref{3.2.14} and then relabelled $\lambda$ with $B_0$. This observation allows us to deduce from \eqref{3.2.15} that (with $\xi^\alpha=(\Psi,\Psi^\dagger,B_i,\pi^i,B_0)$ and $\xi^\beta=(\Psi^\prime,\Psi^{\prime\dagger},B^{\prime}_{j},\pi^{\prime j},B^{\prime}_{0})$)
\begin{equation}
F_{\xi^\alpha\xi^\beta}=\left(
\begin{array}{ccccc}
0&-i&0&0&-e\Psi^{\prime\dagger}\\
i&0&0&0&-e\Psi^\prime\\
0&0&0&-\delta^{i}_{j}&0\\
0&0&\delta^{j}_{i}&0&\partial^{\prime}_{i}\\
e\Psi^\dagger&e\Psi&0&-\partial_j&0
\end{array} \right)\delta({\mathbf x},{\mathbf x}^\prime).\label{3.2.20}
\end{equation}
(Of course this can be calculated directly from the canonical one-form in \eqref{3.2.19a}--\eqref{3.2.19e}.)

Because $F_{\xi^\alpha\xi^\beta}$ in \eqref{3.2.20} is odd dimensional and antisymmetric its determinant must vanish. This means that there is a zero mode present. It is straightforward to show that the zero mode has components given by
\begin{equation}
z^{\xi^\alpha}=(-ie\Psi\theta,ie\Psi^\dagger\theta,\partial_i\theta,0,\theta),\label{3.2.21}
\end{equation}
where $\theta$ is an arbitrary function. It is now possible to show that the field theory analogue of the consistency condition \eqref{2.6} is satisfied identically meaning that there are no new constraints. The symplectic two-form remains degenerate without further conditions. It is easy to understand why this happens. The massless vector field theory that we have written down has a local gauge symmetry and the zero mode found in \eqref{3.2.21} is recognized as the infinitesimal form of the local gauge transformation of both the Schr\"{o}dinger field and the spatial components of the vector field $B_i$. (See \eqref{3.2.4a}--\eqref{3.2.4c}.) The fact that no new constraints are generated and that the consistency conditions are automatically satisfied is just the expression of this gauge symmetry. This is a particularization of the more general treatment given in \cite{montani1993faddeev,montani1993symplectic}. This situation also occurs even if the Darboux transformation has been performed as in the original Faddeev-Jackiw method~\citep{FaddeevJackiw,Jackiwnotears}. It is necessary, as usual, to adopt a gauge condition to remove the gauge degrees of freedom from the theory. The convenient choice here is to pick the Coulomb gauge $\partial^i B_i=0$ and to enforce it by adding on a new Lagrange multiplier field $\lambda$:
\begin{equation}
{\mathcal L}=\pi^i\dot{B}_i+\frac{i}{2}\Psi^\dagger\dot{\Psi}-\frac{i}{2}\dot{\Psi}^\dagger\Psi +B_0(\partial_i\pi^i-e\Psi^\dagger\Psi)+\lambda\,\partial^i B_i+{\mathcal L}^{\prime}_v,\label{3.2.22}
\end{equation}
with ${\mathcal L}^{\prime}_v$ defined in \eqref{3.2.18}. The procedure outlined above now extends the canonical variables to $\xi^\alpha=(\Psi,\Psi^\dagger,B_i,\pi^i,B_0,\lambda)$. In addition to \eqref{3.2.19a}--\eqref{3.2.19e} the canonical one-form picks up a new component
\begin{equation}
A_\lambda=\partial^i B_i.\label{3.2.23}
\end{equation}
This gives some additional components to the symplectic two-form which is now
\begin{equation}
F_{\xi^\alpha\xi^\beta}=\left(
\begin{array}{cccccc}
0&-i&0&0&-e\Psi^{\prime\dagger}&0\\
i&0&0&0&-e\Psi^\prime&0\\
0&0&0&-\delta^{i}_{j}&0&\partial^{\prime i}\\
0&0&\delta^{j}_{i}&0&\partial^{\prime}_{i}&0\\
e\Psi^\dagger&e\Psi&0&-\partial_j&0&0\\
0&0&-\partial^j&0&0&0\\
\end{array} \right)\delta({\mathbf x},{\mathbf x}^\prime).\label{3.2.24}
\end{equation}
There are now no zero modes present and the determinant does not vanish. Making use of \eqref{3.1.14} shows after a bit of calculation that
\begin{equation}
(\det F_{\xi^\alpha\xi^\beta})^{1/2}=\det\lbrack-\nabla^2\delta({\mathbf x},{\mathbf x}^\prime)\rbrack.\label{3.2.25}
\end{equation}
When this result is used in the path integral measure \eqref{2.14} and the field $B_0$ integrated out we recover the form for the path integral that was found by Faddeev~\citep{Faddeev1970} using the Dirac analysis. If we further integrate out the fields $\pi^i$ the standard configuration space path integral result for the Coulomb gauge is found. (See section 12-2-2 of \cite{IZ} for example.)

\section{Discussion}

We have shown how the natural geometrical structure of the Faddeev-Jackiw~\citep{FaddeevJackiw} method gives rise to a measure factor in the path integral for a constrained system. The measure, given by \eqref{2.14} involves the symplectic two-form in a simple way. This extends the analysis of \citep{liao2007path} where the measure required knowing the Jacobian that was involved in the Darboux transformation. This was demonstrated to agree with the path integral found from the more standard Dirac constraint analysis in the examples of Sec.~\ref{Examples}.

Whether one finds the Faddeev-Jackiw, the Dirac, or some other method for dealing with constraints the most suitable is a matter of taste. There are several reasons why we prefer the Faddeev-Jackiw method, in the form advocated in \citep{barcelos1992faddeev}. There does not need to be any classification of constraints into primary and secondary, or first and second class; all can be treated the same. There is no need for weak and strong equalities to hold. Note that in common with \citep{barcelos1992faddeev} we do not advocate solving the constraints to eliminate coordinates or fields, although this does shorten work in some cases. The path integral measure involves the symplectic two-form in a simple way; as this object is central to the Faddeev-Jackiw method no extra work is required, other than calculating its determinant. In the canonical quantization the symplectic two-form needs to be inverted to obtain the canonical commutation relations which is a bit more involved. The path integral measure reproduces both the Faddeev~\citep{Faddeev1970} path integral for first class Dirac constraints, and the Senjanovic~\citep{Senjanovic} path integral for second class Dirac constraints. (As just mentioned we do not need to distinguish between these in the Faddeev-Jackiw method.)

There have been at least two papers that are critical of the symplectic method that we have described here~\citep{RotheandRotheJPA2003,ShirzadandMojiri2005}. Both critiques stem from the fact that the constraints found by the Faddeev-Jackiw method do not necessarily reproduce all of those found by the Dirac method. For example, in the case of the constrained particle on a surface of Sec.~\ref{particle} we found two constraints with a direct physical interpretation. In the Dirac case because there is no time derivative of the coordinate $\sigma$ in \eqref{3.1.2} this leads to a primary constraint in which the canonically conjugate momentum must vanish. The two constraints present in the analysis above are also found. Consistency of these three constraints generates a fourth constraint that also involves the momentum canonically conjugate to $\sigma$. So two constraints found from the Dirac analysis appear to have been missed by the symplectic Faddeev-Jackiw method. This leads \citep{RotheandRotheJPA2003,ShirzadandMojiri2005} to suggest that the symplectic method has a shortcoming and needs either to be modified or else another method must be used. We disagree with this conclusion. First of all there is a fundamental difference between the Dirac and symplectic methods; the Dirac method is a Hamiltonian approach whereas the symplectic approach is a Lagrangian method. There is no reason to believe that there must be a one-to-one correspondence between the constraints for the two methods. The two extra constraints generated by the Dirac method for the constrained particle are simply because the method is a Hamiltonian one; furthermore, unlike the two constraints that we found above have no direct physical meaning. In fact in his Dirac path integral treatment Kashiwa~\citep{kashiwa1996general} simply drops them. However what must be true is that both the Dirac method and the symplectic Faddeev-Jackiw method must yield the same physics. The examples of Sec.~\ref{Examples} demonstrate that with the choice of measure adopted in this paper both methods result in the same path integral and hence the same physical consequences. There is no need for modifications of the method.

%\bibliography{FJPI}
%\bibliographystyle{apsrev}

\end{document}